\documentstyle[psfig,aps,twocolumn]{revtex}
\begin{document}
\title{Quasiclassical theory of twin boundaries in High-$\mbox{T}_{\text{c}}$
 superconductors}
\author{W. Belzig$^1$, C. Bruder$^1$, and M. Sigrist$^2$}
\address{$^1$ Institut f{\"u}r Theoretische Festk{\"o}rperphysik, Universit{\"a}t
Karlsruhe, D-76128 Karlsruhe, Germany\\
$^2$ Yukawa Institute for Theoretical Physics, Kyoto University, Kyoto 606-01,
Japan}
\address{}
\date{\today}
\maketitle
\begin{abstract}
We investigate the electronic structure of twin boundaries in
orthorhombically distorted high-T$_c$ materials using the
quasiclassical theory of superconductivity. At low temperatures we
find a local instability to a time-reversal symmetry breaking state at
the twin boundary. This state yields spontaneous currents along the
twin boundary that are microscopically explained by the structure of
the quasiparticle bound states. We calculate the local density of
states and find a splitting in the zero-bias peak once the spontaneous
currents set in. This splitting is measurable by STM techniques and is
a unique signature of time-reversal breaking states at a twin
boundary.
\end{abstract}

A large number of experiments have convincingly demonstrated that
Cooper pairs have basically $ d_{x^2-y^2} $-wave symmetry in
high-temperature superconductors \cite{vanharlingen}. In a strict
sense this classification applies only to superconductors with
perfectly tetragonal crystal symmetry where the $ d_{x^2-y^2} $
represents a ``relative angular momentum'' of the Cooper pairs with
reduced symmetry. On the other hand, there is a number of slightly
orthorhombically distorted systems such as YBa$_2$Cu$_3$O$_7$ (YBCO)
where the intrinsic crystal deformation removes the symmetrical
distinction between this pairing state and a conventional $s$-wave
type. We may interpret this also in the way that the orthorhombic
distortion couples the two pairing channels
\cite{sigrist,walker}. Clear evidence for this kind of mixing has been
found in recent c-axis Josephson experiments between YBCO and
Pb\cite{kouznetsov}. These experiments established the presence of an
$s$-wave component and the dominance of the $ d_{x^2-y^2} $-wave
symmetry.

An interesting aspect of the $s$-wave admixture due to orthorhombic
distortion occurs in the vicinity of twin boundaries (TB) which
separate the two degenerate orthorhombic lattice shapes (twins). It
was suggested that the two order parameter components, $s$- and
$d$-wave, could locally twist in a way that time reversal-symmetry $
{\cal T} $ is broken. Various physical properties are connected with
this effect. In Ref.\cite{sigrist} it was shown on the level of
Ginzburg-Landau theory that the $ {\cal T} $-violating state is
accompanied by a spontaneous current flowing parallel to the twin
boundary. Recently this system was studied by solving the Bogolubov-de
Gennes (BdG) equations \cite{feder,zhitomirsky} confirming the results
of the Ginzburg-Landau description and discussing the local $ IV
$-characteristics observable in a tunneling experiment. Time-reversal
breaking surface states were also predicted \cite{fogelstrom} and
experimentally found \cite{greene} at surfaces of YBCO.

In this paper, we study this type of system and its properties using
the quasiclassical theory of superconductivity developed by
Eilenberger \cite{rainer}. We calculate the structure of the order
parameter self-consistently at all temperatures. Our calculation
includes the orthorhombic distortion in form of anisotropic
quasiparticle masses. In contrast to Ref. \cite{feder}, the twin
boundary has no extension and is not explicitly pair breaking. In the
bulk this formulation gives a mixing between $s$- and $d$-wave pairing
due to the orthorhombic distortion in agreement with the other
methods. For the twin boundary we show that the time-reversal symmetry
is indeed broken at low enough temperature. The lower symmetry is
noticeable also in the energy levels of quasiparticle bound states at
the TB. In the $\cal{T}$-invariant state there is a large density of
bound states at zero-energy. The breakdown of time-reversal symmetry
in the TB leads to the splitting of this energy level, and this effect
in the local density of states is observable by STM (scanning
tunneling microscopy). The properties of these bound states lead also
to a microscopic interpretation of the spontaneous current along the TB.

In YBCO the orthorhombic symmetry reduction is partially induced by
lattice deformation and mainly due to the presence of CuO-chains. 
The introduction of this orthorhombicity into the quasiclassical
formulation is done by changing the single electron dispersion to
\begin{equation}
 \label{dispersion}
 \varepsilon(\bbox{k})=\frac{\bbox{k}^2}{2m}-\frac{c}{m}k_x k_y\; ,
\end{equation}
where the (dimensionless) constant $-1< c<1$ parameterizes the
distortion (the sign of $c$ defines the two twin domains). Here the
$x$- and $y$-axes correspond to the direction 
[1,1,0] and [1,-1,0] of the tetragonal lattice. 
This leads to a non-cylindrical FS with $ 
 k_{\text{F}}(\theta)=\frac{k_{\text{F0}}}{\sqrt{1-c\sin 2\theta}} $
where $\theta$ is the angle between $\bbox{k}$ and the $k_x$-direction
and $k_{\text{F0}}$ the Fermi wave vector of the undistorted system.
We will neglect all effects of electron motion in $c$-axis direction.
With this anisotropy the averaging over the FS has to be modified
compared to the standard case,
\begin{equation}
 \label{average} \sum_{\bbox k} \to N_0\langle \cdots\rangle
 =N_0\int_0^{2\pi}\frac{d\theta}{2\pi} n(\theta)\; .
\end{equation}
Here, $N_0$ is the normal density of states at the Fermi energy, and
\begin{equation}
 \label{ndef}
 n(\theta)=n_{\text{f}}\frac{\bbox{k}_{\text{F}}^2(\theta)}
 {m|\bbox{k}_{\text{F}}(\theta) 
 \bbox{v}_{\text{F}}(\theta)|} = n_{\text{f}}\frac{1}{1-c\sin 2 \theta}\;,
\end{equation}
where $n_{\text{f}}$ is a normalization factor. 
A realistic estimate for the parameter $ c $ is $ \sim 0.2 $; this is
mainly due to the chains \cite{atkinson}. Besides this
modification of the dispersion we will neglect all other effects
related to the chains. 

In the following, we restrict our discussion to the
$s$-wave and $d_{x^2-y^2}$-wave pairing channels.
The pairing interaction is modeled as
\begin{equation}
 \label{pairing}
 V(\theta,\theta^\prime)=V_{\text{s}}+2V_{\text{d}}
 \sin(2\theta)\sin(2\theta^\prime)\; , 
\end{equation}
where the angular dependence of the $d$-wave part follows from our
choice of the coordinate frame. 
The coupling strengths $V_{\text{s/d}}$ are eliminated in favor of the bare
critical temperatures $T_{\text{cs/d}}$ and the energy cut-off
$\omega_{\text{c}}$, the latter assumed to be the same for both channels. Then
the coupling constants can be defined as $V_{\text{s,d}}^{-1}=
\ln(T/T_{\text{c,d}}) + \sum_{n>0}^{n<\omega_{\text{c}}/2\pi T}(n+1/2)^{-1}$.

The bulk solution for the off-diagonal quasiclassical Green's function is
$f_\omega(\theta) = \Delta(\theta)/\Omega(\theta)$, where
$\Omega(\theta)=\sqrt{\omega^2+|\Delta(\theta)|^2}$. 
The gap function is a linear combination of the $s$-
and $d$-wave component, $\Delta(\theta) = \Delta_d
\sin(2\theta)+\Delta_s$. 

\begin{figure}[tbp]
 \begin{center}
 \leavevmode
 \psfig{figure=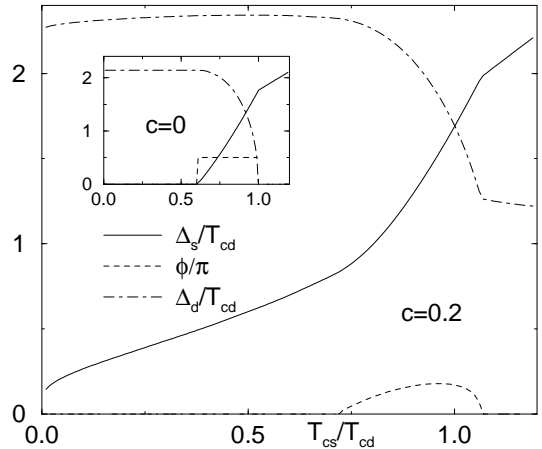,width=7cm}
 \caption{Zero-temperature properties of an orthorhombic
 superconductor with mixed $d$- and $s$-pairing. For different
 magnitudes of the orthorhombic distortion $c=0.2$ (main graph) and
 $c=0$ (inset), the two order parameters and the relative phases are
 shown as a function of $T_{cs}/T_{cd}$.}
 \label{fig:delta.bulk}
 \end{center}
\end{figure}

The self-consistent calculation leads to
two coupled equations for the $s$- and $d$-wave component. In the
case $c=0$ we either find a one-component solution (i.e., the $s$- or the
$d$-wave component vanishes), or both are finite and
appear in the complex combination $d\pm is$,
i.e. they have the relative phase $\phi = \pm \pi/2$
\cite{matsumoto}. At $T=0$ the complex phase occurs for
$0.6\lesssim T_{cs}/T_{cd} < 1.0$ (see the inset in
Fig.~\ref{fig:delta.bulk}).

An expansion in $c$ immediately shows that the $s$ and $d$ components
are coupled for $c\neq 0$ and a finite value of one of the components
drives the other component to be non-zero. A non-vanishing $c$ also leads
to a renormalized onset temperature for superconductivity, 
\begin{equation}
 \label{eq:crittemp}
 T_c(c)=\frac{1}{2}(T_{cs}+T_{cd})+\frac{1}{2}
 \sqrt{(T_{cs}-T_{cd})^2+2\tilde{c}^2T_{cs}T_{cd}}\; ,
\end{equation}
where $\tilde{c}=c\sum_{n=0}^{\omega_c/2\pi T_c}1/(n+1/2)$. 

The numerical results for the two order parameters and their relative
phase at $T=0$ are shown in Fig.~\ref{fig:delta.bulk} as a function of
$T_{cs}/T_{cd}$.  They always coexist. In contrast to the case $c=0$,
the relative phase $\phi$ varies continuously and is different
from zero in a narrow window of values of $ T_{cs}/T_{cd} $.  The
region of broken time-reversal symmetry is shrinking with increasing
$|c|$, in qualitative agreement with the GL approach.

In the following we will assume $T_{cs}/T_{cd} = 0.3 $ (see
Ref.~\cite{fogelstrom}) and $c=0.2$ which leads to $\Delta_s \sim 0.25
\Delta_d$ and $ \phi =0 $ or $ \pi $ (see
Fig.~\ref{fig:delta.bulk}), i.e., no bulk $\cal{T}$-violating state
occurs in agreement with the present experimental status.

We will now consider a TB, a boundary between two domains with the
relative orientation of the crystal axes of $90^{\circ}$. The
quasiclassical equations have to be solved along classical
trajectories, which are characterized by a momentum direction on the
Fermi surface $\bbox{k}_{\text{F}}$ \cite{rainer}. At the TB the
k-vectors with the same $k_y$-component have to be matched. The
reflection coefficient for a given $\bbox{k}_{\text{F}}$ in the
absence of a boundary potential is
$R=(v_{\text{Fx1}}-v_{\text{Fx2}})^2/(v_{\text{Fx1}}+v_{\text{Fx2}})^2$.
If we consider the energy dispersion given in Eq.~(\ref{dispersion})
($ c = \mbox{sign}(x) |c| $), we find that
$v_{\text{Fx1}}=v_{\text{Fx2}}$ for all scattering
directions. Consequently we have {\em no normal} reflection at the TB
\cite{zhitomirsky}.

\begin{figure}[htbp]
 \begin{center} 
 \leavevmode 
 \psfig{figure=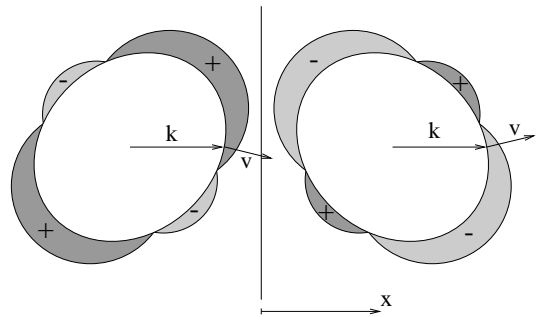,width=7cm}
 \vspace{0.2cm}
 \caption{Fermi surfaces and bulk gap (shaded region) on both sides of
 the TB. Fermi wave vectors $\bbox{k}_F$-vectors and
 Fermi velocities $\bbox{v}_F$ for a perpendicular trajectory are
 indicated by arrows. The parallel components of $v_F$ for these
 trajectories on both sides of the TB have different sign.}
 \label{fig:twinbound}
 \end{center}
\end{figure}

We solve the quasiclassical equations numerically in a self-consistent
way for various temperatures for a single twin boundary. The
calculation was simplified by applying the Schopohl-Maki
transformation to the Eilenberger equations \cite{schopohl}. 
Consistent with experimental observations \cite{vanharlingen,kouznetsov}
we assumed that the $d$-wave component is identical on both sides, whereas
the $s$-wave component changes sign.

\begin{figure}[tbp]
 \begin{center} 
 \leavevmode 
 \psfig{figure=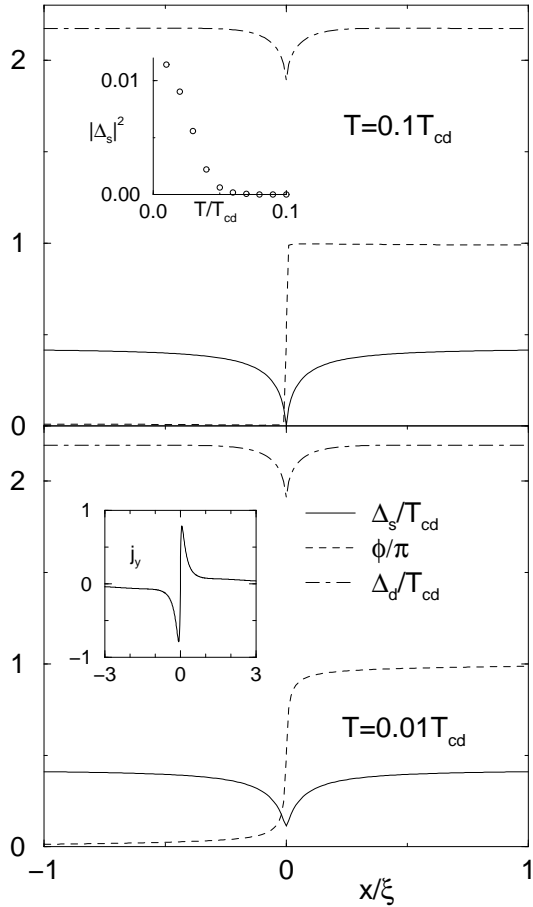,width=7cm}
 \caption[]{Order parameters and relative phase in the vicinity of a
 twin boundary for $T_{cs}/T_{cd}=0.3$ and $c=0.2$. Upper graph:
 $\cal{T}$-invariant state of the TB for $T=0.1T_{cd}$. The $s$-component
 goes to zero at the TB. Lower graph:
 locally $\cal{T}$-violating state for $T=0.01T_{cd}$. The
 absolute value of the $s$-component is finite for $x=0$ and the
 relative phase changes smoothly from $0$ to $\pi$. Inset upper graph:
 Modulus squared of (imaginary) $s$-component as a function of temperature.
 Inset lower graph: spontaneous current density in units of
 $10^{-1}eN_0T_{cd}v_F$.}
 \label{fig:selftwin} 
 \end{center}
\end{figure}

Results for two different temperatures are shown in
Fig.~\ref{fig:selftwin}. For the higher temperature $T=0.1 T_{cd} $
the relative phase jumps from $0$ to $\pi$ at the twin boundary and
$\Delta_s$ goes to zero directly at the TB. On the other hand, for
$T=0.01T_{cd}$ we observe a local $\cal{T}$-violation where the
relative phase changes continuously from $0$ to $\pi$ with a
non-vanishing $s$-wave pairing component. The inset in the upper graph
shows $|\Delta_s|^2$ at the TB as a function of temperature. It
becomes finite for $T\lesssim 0.05$. A particular feature of local
$\cal{T}$-violation is the appearance of spontaneous currents in the
vicinity of the TB flowing in opposite directions on both sides.
These spontaneous currents at $T=0.01 T_{cd}$ are shown in
Fig.~\ref{fig:selftwin} (inset lower graph). Note that screening
effects are not taken into account here.

\begin{figure}[htbp]
 \begin{center} 
 \leavevmode 
 \psfig{figure=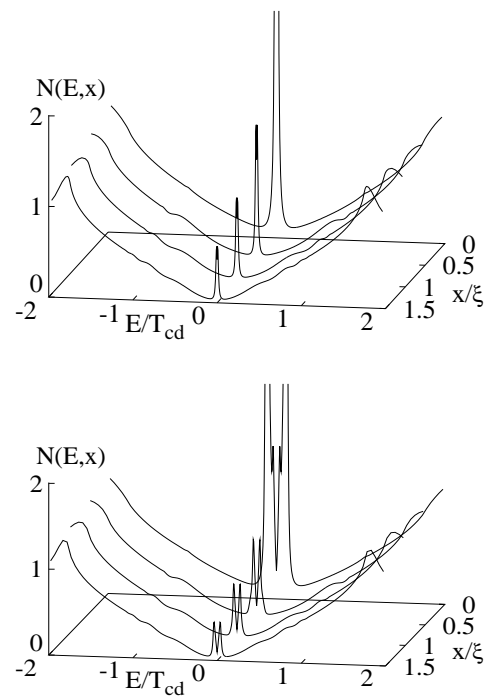,width=9cm}
 \caption[]{Local density of states in the vicinity of a twin boundary
 for $T_{cs}/T_{cd}=0.3$ and $c=0.2$. The two graphs correspond to
 $T=0.1T_{cd}$ (upper graph) and $T=0.01T_{cd}$ (lower graph). In the
 $\cal{T}$-invariant state there is a large peak at zero energy. The
 peak splits symmetrically in the $\cal{T}$-violating state.}
 \label{fig:twindos}
 \end{center}
\end{figure}

The local density of states (LDOS) can be calculated by solving the
Eilenberger equation for real energies. The results are given in
Fig.~\ref{fig:twindos}. In the $\cal{T}$-invariant state a pronounced
zero-energy peak occurs in the vicinity of the TB. In the
$\cal{T}$-violating state, the zero-energy peak splits into two
symmetric peaks. The separation of the peaks is proportional to
$\Delta_s(0)$. The LDOS can be measured by scanning tunneling
microscopy (STM). In particular, the splitting of the zero-energy peak
would prove the existence of a time-reversal breaking state in YBCO
\cite{feder}. The temperature at which the splitting occurs, $T \sim
0.05T_{cd} \sim O(5K)$, is well within reach of current
low-temperature STM technology.

To investigate the microscopic nature of the gap feature in the LDOS
and the spontaneous current it is useful to turn to an angle-resolved
view of the problem. The zero-energy peak in the LDOS in the
$\cal{T}$-invariant situation can be understood by considering
classical trajectories with angles $ \theta $ centered around 0, 
$\pi$ and $\pm \pi/2 $ in a range $ 2 \delta \theta = {\rm arcsin} (
\Delta_s/ \Delta_d ) $. The corresponding Andreev (electron-hole)
bound state for such angles is confined between two superconducting
domains with a phase shift of $ \pi $ which leads to a zero-energy
level \cite{hu}. On the other hand, broken time-reversal symmetry will
generate phase shifts different from $ \pi $ leading to up or down
shifted energy levels and the splitting of the zero-energy
peak. Alternatively, this effect may be also seen as a driving
mechanism: the large DOS of the zero-energy bound states at the Fermi
energy gives rise to a local Fermi surface instability opening a
(pseudo) gap.

How does this splitting lead to spontaneous currents? Each state
(bound or extended) carries a current parallel to
$\bbox{v}_F(\bbox{k})$. The phase gradient seen by these states leads
to a shift in energy (bound states) or spectral weight (extended
states). As a result left- and right-moving states are differently
occupied. At $T=0$ only the {\em right}-moving of the bound states
are occupied. On the other hand, due to the shift of the spectral
weight the {\em left}-moving continuum states give the dominant
contribution of the continuum states. They cancel the perpendicular
current contribution of the bound states (as required by current
conservation), leaving only a parallel component of the total
current\cite{huck}. The sign of the parallel current is different on
both sides of the TB due to the symmetry of the dispersion relation
(the momenta and velocities for one of these states are indicated in
Fig.~\ref{fig:twinbound}).

In conclusion, we have studied the electronic structure of twin
boundaries in orthorhombically distorted high-T$_c$ superconductors.
The orthorhombic distortion was introduced in the quasiclassical
formalism of superconductivity, and the corresponding equations were
solved self-consistently. In contrast to Ref.~\cite{feder} we do not
assume that the TB has a pair-breaking effect. At low temperatures,
we found a localized ${\cal T }$-violating state at the TB. These
localized states create spontaneous quasiparticle currents that flow
in parallel to the TB. In addition, we find a splitting of the
zero-bias anomaly in the local density of states. Using realistic
parameters we estimate that this splitting could be observed around
$5$K, i.e., using available low-temperature scanning tunneling
microscopes. This splitting would be a unique signature of
$\cal{T}$-violation in such systems.

We would like to acknowledge helpful discussions with A. van Otterlo,
K. Kuboki, T.M. Rice, D. Agterberg, and, in particular, Yu.~Barash.
W.~B. and C.~B. were supported by DFG grant No.~Br1424/2-1.


\begin{thebibliography}{99}
\bibitem{vanharlingen} D.~J. van Harlingen, Rev. Mod. Phys. {\bf 67}, 515 (1995).
\bibitem{sigrist} M. Sigrist, K. Kuboki, P.~A. Lee, A.~J. Millis, and
T.~M. Rice, Phys. Rev. B {\bf 53}, 2835 (1996).
\bibitem{walker} M.~B. Walker, Phys. Rev. B {\bf 53}, 5835 (1996);
M.~B. Walker and J. Luettmer-Strathmann, Phys. Rev. B {\bf 54}, 588 (1996).
\bibitem{kouznetsov} K.~A. Kouznetsov {\it et al.}, Phys. Rev. Lett. {\bf 79},
3050 (1997).
\bibitem{feder} D.~L. Feder, A. Beardsall, A.~J. Berlinsky, and C. Kallin,
Phys. Rev. B {\bf 56}, R5751 (1997).
\bibitem{zhitomirsky} M.~E. Zhitomirsky and M.~B. Walker, Phys. Rev. Lett. 
{\bf 79}, 1734 (1997); M.~E. Zhitomirsky and M.~B. Walker, preprint
cond-mat/9709282.
\bibitem{fogelstrom} M. Fogelstr\"om, D. Rainer, and J.~A. Sauls,
Phys. Rev. Lett. {\bf 79}, 281 (1997).
\bibitem{greene} M. Covington {\it et al.}, Phys. Rev. Lett. {\bf 79}, 277 
(1997).
\bibitem{rainer} G. Eilenberger, Z. Physik {\bf 214}, 195 (1968). 
For a comprehensive review of the quasiclassical formalism 
applied to unconventional superconductors, see the forthcoming 
{\it Quasiclassical Methods in Superconductivity and Superfluidity},
edited by D. Rainer and J.~A. Sauls (Springer, Heidelberg, 1998).
\bibitem{atkinson} W.~A. Atkinson and J.~P. Carbotte, Phys. Rev. B
{\bf 52}, 10601 (1995).
\bibitem{matsumoto} M. Matsumoto and H. Shiba, J. Phys. Soc. Jpn. {\bf 64},
3384 (1995).
\bibitem{schopohl} N. Schopohl and K. Maki, Phys. Rev. B {\bf 52}, 490
(1995).
\bibitem{hu} C.-R. Hu, Phys. Rev. Lett. {\bf 72}, 1526 (1994);
Y. Tanaka and S. Kashiwaya, Phys. Rev. B {\bf 53}, 11957 (1996).
\bibitem{huck} For a simpler situation with a related mechanism see
A. Huck, A. van Otterlo, and M. Sigrist, Phys. Rev. B {\bf 56}, 14163 (1997).
\end{thebibliography}
\end{document}